\documentclass[preprint,12pt]{aastex}
\usepackage{epsfig, amsmath, amsfonts}

\newcommand\bb[1] {   \mbox{\boldmath{$#1$}}  }

\newcommand\del{\bb{\nabla}}
\newcommand\bcdot{\bb{\cdot}}
\newcommand\btimes{\bb{\times}}

\newcommand{\BV}{Brunt-V\"ais\"al\"a\ }

\newcommand{\dd}[2]{\frac{{\rm d} #1}{{\rm d} #2}}

\def\dd{\partial}

\def\beq{ \begin{equation} }
\def\eeq{ \end{equation} }
\def\spose#1{\hbox to 0pt{#1\hss}}  
\def\ltsim{\mathrel{\spose{\lower.5ex\hbox{$\mathchar"218$}}
\raise.4ex\hbox{$\mathchar"13C$}}}
\def\gtsim{\mathrel{\spose{\lower.5ex\hbox{$\mathchar"218$}}
\raise.4ex\hbox{$>$}}}

\begin{document}

\title{\bf\LARGE A Simple Model for Solar Isorotational Contours}
\author{ Steven A. Balbus\altaffilmark{1,2}}

\altaffiltext{1}{Laboratoire de Radioastronomie, \'Ecole Normale
Sup\'erieure, 24 rue Lhomond, 75231 Paris CEDEX 05, France
  \texttt{steven.balbus@lra.ens.fr}}

  \altaffiltext{2}{Adjunct Professor, Department of Astronomy, University of Virginia,
  Charlottesville VA 22903}

	\begin{abstract}

The solar convective zone, or SCZ, is nearly adiabatic and marginally
convectively unstable.  But the SCZ is also in a state of differential
rotation, and its dynamical stability properties are those of a weakly
magnetized gas.  This renders it far more prone to rapidly growing
rotational baroclinic instabilities than a hydrodynamical system would be.
These instabilities should be treated on the same footing as convective
instabilites.  If isentropic and isorotational surfaces coincide in
the SCZ, the gas is marginally (un)stable to {\em both} convective
and rotational disturbances.  This is a plausible resolution for the
instabilities associated with these more general rotating convective
systems.  This motivates an analysis of the thermal wind equation in which
isentropes and isorotational surfaces are identical.  The characteristics
of this partial differential equation correspond to isorotation contours,
and their form may be deduced even without precise knowledge of how
the entropy and rotation are functionally related.  Although the exact
solution of the global SCZ problem in principle requires this knowledge,
even the simplest models produce striking results in broad agreement with
helioseismology data.  This includes horizontal (i.e. quasi-spherical)
isorotational contours at the poles, axial contours at the equator,
and approximately radial contours at midlatitudes.  The theory does not
apply directly to the tachocline, where a simple thermal wind balance is
not expected to be valid.  The work presented here is subject to tests
of self-consistency, among them the prediction that there should be good
agreement between isentropes and isorotational contours in sufficiently
well-resolved large scale numerical MHD simulations.

\end{abstract}


\maketitle




\section{Introduction}

The problem of differential rotation in the solar convective zone
(SCZ) has vexed theorists for many years.  At the simplest level, one
is faced with the fact that the only region of the sun characterized by
significant turbulence is also the only region showing any significant
differential rotation.  This is a cogent reminder of the hazards of
modeling a rotating, turbulent gas by an effective viscosity parameter.
The SCZ is far too complex to yield to such an approach.  Indeed,
large scale numerical simulations have been in place for many years,
trying to elicit the rotation profile of the sun's
outer layers (Brummell, Cattaneo, \& Toomre 1995; Thompson et al. 2003).
While these studies are now at a stage where they can begin to make
contact with the observational data, they have not yet been able to
reproduce the salient features of the observed solar rotation profile.
This is especially true for the isorotational
contours (hereafter ``isotachs''), which tend to be cylindrical
in simulations, but are decidedly noncylindrical in the sun.

In this paper, we examine an effect that could be important for
understanding the SCZ rotation profile, but has not been sufficiently
emphasized in previous investigations: the extreme sensitivity of
ionized differentially rotating gas to the presence of even very
weak magnetic fields of arbitrary geometry (Balbus \& Hawley 1994,
Ogilvie 2007).  Far from behaving as a passive vector field, a weak
magnetic field triggers rapidly growing unstable local modes in rotating
systems that would be hydrodynamically unstable.  The field endows the
gas with degrees of freedom that have no hydrodynamical counterpart.
The classical manifestation of this behavior takes the form of what
is known as the magnetorotational instability (MRI), widely regarded
as the process responsible for turbulence and enhanced transport in
accretion disks (Balbus \& Hawley 1991, 1998).   But destabilization
by a weak magnetic field is a process that extends beyond the domain
of rotationally supported disks.  Such fields are agents of dynamical
destabilization in baroclinic and convective flows as well (Balbus 1995),
with potentially interesting applications to the SCZ.

It is essential to understand two key counterintuitive points
at the outset.  The first is that destabilization by a magnetic
field can occur even when the field plays essentially no role in the
dynamical equilibrium: a weakly magnetized gas does not behave ``almost
hydrodynamically'' with respect to its stability properties.  The second
is that the destabilization is independent of the field strength and
field geometry.  In MRI simulations, the disruption remains vigorous and
self-sustaining even in highly convoluted turbulent flow.  This is what
makes MHD processes potentially so important for understanding rotating,
stratified flows.

It is the emphasis on these points that sets the present approach
apart from earlier work that also addressed the stability of rotating,
stratified, magnetized flows (e.g. Acheson 1983, Ogden \& Fearn
1995).  In these earlier studies, the focus is upon toroidal fields,
nonaxisymmetric disturbances, and instability driven directly by the
magnetic field.  By contrast, what is crucial here are poloidal fields
and axisymmetric disturbances.  Moreover, while the instability of
interest relies on the presence of a magnetic field, the (nonconvective)
seat of free energy is the differential rotation, not the field itself.
The role of the field is to provide the critical degrees of freedom to
the fluid required to tap into the (destabilizing) differential rotation.

We shall begin our presentation, however, not with magnetic fields,
but with a theoretical observation that may be taken at a purely
phenomenological level, independently of deeper MHD considerations.
This is the remarkable agreement with the overall pattern of solar
isorotation contours produced by a simple calculation in which a dominant
thermal wind balance from the vorticity equation is combined with the
assumption that entropy and angular velocity gradients are counteraligned.
Whatever the cause of the counteralignement may be, the analysis on its
own is highly suggestive.  (A connection between entropy and angular
velocity gradients has in fact been advocated on purely hydrodynamical
grounds [Miesch, Brun, \& Toomre 2006].)  This material is presented in
\S 2.  In \S 3, we discuss in some detail the case for an MHD coupling
between angular velocity and entropy gradients.  The final section is
a critical discussion of unresolved issues provoked by this work, and
a summary.


\section {Solving the thermal wind equation}

\subsection {Review}\label{revsec}

Let $(R, \phi, z)$ be a standard cylindrical coordinate system, and $(r,
\theta, \phi)$ a standard spherical coordinate system.  Unit vectors
are denoted $\bb{e_R}$, $\bb{e_\theta}$, etc.  The angular
velocity $\Omega$ is assumed to be independent of $\phi$, but
otherwise general.  Our notation for the fluid variables is likewise
standard: $\bb{v}$ is the velocity, $P$ is the gas pressure, $\rho$
is the mass density, and $\bb{B}$ is the magnetic field.

We adopt a fiducial value of $2.5\times 10^{-6}$ rad s$^{-1}$
for the angular velocity $\Omega$ in the SCZ, corresponding to rotation
velocities between 1 and 2 km s$^{-1}$.  This is well in excess of the
$\sim 30$ m s$^{-1}$ expected of convective velocities, but such a direct
comparison is not necessarily the most relevant one.  
A more telling comparison is between the squared \BV frequency
\beq\label{fiduN}
\left| N^2\right|  = \left| -{1\over \rho\gamma}{\dd P\over \dd r} {\dd\ln P\rho^{-\gamma}
\over \dd r}\right| \sim 3.8\times 10^{-13}\ {\rm s}^{-2}
\eeq
(we have adopted a value of $10^{-6}$ for $\dd\ln P\rho^{-\gamma}/\dd\ln r$
[Schwarzschild 1958])
and the rotational parameter
\beq\label{fiduOm}
{d\Omega^2\over d\ln R} \sim 1.8\times 10^{-12} \ {\rm s}^{-2}.
\eeq
This is evidently about 5 times as large as $N^2$. (Whether $R$
or $z$ is used in the angular velocity gradient is not critical at
midlatitudes.)  Taking this estimate at face value, the rotational
parameter is significantly larger than $N^2$.  The significance of this
will shortly become apparent.

Let us first consider systems whose
equilibrium velocity is differential rotation, $\bb{v} = R\Omega(R,
z)\bb{e_\phi}$.
The thermal wind equation follows directly
from the time-steady form of the
vorticity equation, 
\beq\label{vort2}
R{\dd\Omega^2\over \dd z} = {1\over \rho^2}\left(
\del P\btimes \del\rho 
\right)\bcdot\bb{e_\phi}.
\eeq
Expressing the right side in $r,\theta,\phi$
spherical coordinates:
\beq
R {\dd\Omega^2\over \dd z} = {1\over r \rho^2}
\left( {\dd\rho\over\dd \theta}{\dd P\over \dd r} -
{\dd\rho\over\dd r}{\dd P\over \dd \theta} 
\right).
\eeq
Now, rewrite this in terms of the entropy gradients:
\beq
R {\dd\Omega^2\over \dd z} = {1\over C_P  \rho r}
\left( {\dd P \over\dd \theta}{\dd S\over \dd r} -
{\dd P \over\dd r}{\dd S\over \dd \theta} 
\right),
\eeq
where $S$ is the specific entropy,
\beq
S={k\over \gamma -1} \ln P\rho^{-\gamma} + {\rm constant},
\eeq
$k$ is the Boltzmann constant, and $C_P$ is the constant
pressure specific heat,
\beq
C_P =\left(\gamma\over \gamma - 1\right) k
\eeq
In the SCZ, the $r$ gradient of $P$ clearly dominates
over the
$\theta$ gradient, whereas the $r$ gradient of $S$ is
unlikely to greatly exceed the $\theta$ gradient.
(In fact, we shall presently argue just the opposite.)
We may then conclude that
\beq\label{thwnd}
R {\dd\Omega^2\over \dd z} = {g\over \ C_P r } {\dd S\over\dd\theta}
={g\over \gamma r } {\dd(\ln P\rho^{-\gamma}) \over  \dd\theta}
\eeq
where $g=-(1/\rho) {\dd P/\dd r}$ is the gravitational field in
hydrostatic equilibrium, ignoring the small effects of rotation.
Significant latitudinal entropy gradients are required to
avoid cylindrical isotachs.

Equation (\ref{thwnd}) is known as the thermal wind equation
(Kitchatinov \& R\"udiger 1995, Thompson et al. 2003).  This equation
holds in our MHD analysis, because we explicitly assume that the magnetic
fields are sufficiently weak in the equilibrium state that they do not
affect the large scale rotation profile.  (For a 10 G field and a
density of
$0.05$ g cm$^{-3}$, the Alfv\'en velocity is 13 cm per second.)
In full spherical coordinates, the thermal wind equation is
\beq\label{thw2}
\left(
\cos\theta{\dd\Omega^2\over \dd r}  -
{\sin\theta\over r} {\dd\Omega^2\over \dd \theta} 
\right)
 = {g\over C_P r^2\sin\theta } {\dd S\over\dd\theta}
\eeq

Even before a detailed solution is developed, a simple scaling argument
reveals something of interest here.  The $r$ gradient of $S$
is generally determined by the need to transport 
the solar luminosity by thermal convection
(Schwarzschild 1958, Clayton 1983).  On the other hand,
the $\theta$ gradient is, from equation (\ref{thwnd}), linked directly
to differential rotation.  If the fiducial numbers (\ref{fiduN})
and (\ref{fiduOm}) are reasonably accurate, at midlatitudes equation
(\ref{thwnd}) implies that the $\theta$ gradient of $S$ will significantly
exceed the $r$ gradient.  The point of interest here is that this
anisotropic feature is directly seen in the $\Omega$ gradient deduced from the
helioseismology data.  This suggests that there is a deeper dynamical
coupling present between $S$ and $\Omega$, beyond just the general trend that
one goes up as the other goes down.  We will develop this idea more
fully in the next two sections.

\subsection{Analysis}\label{anall}

The thermal wind equation is a familiar tool to practitioners of
solar rotation theory.   The gross features of the sun's angular
velocity profile ($\Omega$ decreasing polewards from the equator) can
be understood relatively simply with the aid of this equation and some
reasonable assumptions of the efficiency of convection in the presence
of Coriolis forces (Thompson et al. 2003).  The idea is that convection
in the equatorial direction is impeded by Coriolis forces, resulting in a
more efficient transport of heat along the rotation axis.
The poles are then regions of higher specific entropy compared with the
equatorial zone, and the resulting $\theta$ gradient in $S$ drives axial
gradients in $\Omega$.

This is a well-motivated and physically plausible scenario.  It seems
natural therefore, to search for explicit model solutions to the thermal
wind equation incorporating this idea.  Moving upward from the equator,
we expect the angular velocity to decrease as the entropy increases.
Clearly the gradients of these quantities are in broadly opposite senses.
But the final paragraph of \S \ref{revsec} suggests that ``counteraligned'' may
be a better description then ``broadly opposite.''  This raises the
question of what the solutions to equation (\ref{thw2}) would look like
if the two gradients were in fact {\em precisely} oppositely aligned.

Following the notion that 
isentropic and isorotational surfaces coincide, we assume that
$S=S(\Omega^2)$.
($S$ should not change if $\Omega$
changes sign, hence the dependence on $\Omega^2$.)  
Equation (\ref{thw2}) then becomes
\beq\label{maineq}
{\dd \Omega^2\over \dd r} 
-
\left(
{g S' \over C_P r^2\sin\theta \,\cos\theta}+
{\tan\theta\over r} 
\right)
{\dd \Omega^2 \over\dd\theta} =0,
\eeq
where $S'\equiv dS/d\Omega^2$.
The solution of equation (\ref{maineq}) is that
$\Omega^2$
is constant along the characteristic
\beq\label{char1}
{d\theta\over dr} =  - {\tan\theta\over r} - {g S'\over C_P r^2\sin\theta\ \cos\theta }
\eeq
But if $\Omega^2$ is constant along this characteristic, then so must be $S'$.
This is therefore a self-contained, ordinary differential equation for $\theta$
as a function of $r$, precisely the isorotational contours that we seek.
To solve equation (\ref{char1}), let $y=\sin\theta$.  Then,
our differential equation simplifies to
\beq
{dy^2\over dr} +{2y^2\over r} = - {2g S' \over C_P r^2}
\eeq
Multiplying by $r^2$ and regrouping,
\beq
{d(r^2y^2)\over dr} = - {2g S' \over C_P }= - {2GM_\odot S' \over C_P  r^2 }
\eeq
where $G$ is Newton's constant and $M_\odot$ is a solar mass.
(We have ignored the local self-gravity of the SCZ.)  Since
$S'$ is a constant along the characteristic, this integrates
immediately to
\beq\label{conts}
r^2\sin^2\theta = R^2 = A - {B\over r}
\eeq
where $A$ is a constant of integration and
\beq
B= -  
{2GM_\odot S' \over C_P   }.
\eeq
We have inserted a minus sign since $S$ will generally be a decreasing
function of $\Omega^2$.  Indeed, as will become very clear, the
characteristics make no sense if $S'$ is positive, but a very great
deal of sense if it is negative.  In this model, the solar isotachs
are given by a remarkably simple formula.

To estimate the magnitude of $B$, note that it may
be written
\beq
{B\over r_\odot^3}= \left( 2GM_\odot/r_\odot\over \gamma r_\odot^2\Omega^2\right)
\left( -{d\ln P\rho^{-\gamma}\over d\ln r}\right)
\left({d\ln r\over d\ln \Omega^2}\right)
\eeq
The first factor is large, of order $10^5$.   The second, we have already
estimated at $10^{-6}$.  The helioseismology data suggest that the
third factor ranges between 1 and 10, and is larger near the equator.
Crudely speaking, we expect $B/r_\odot^3$ to be of order unity
or less.  There should not be a large difference in scale
between $A/r_\odot^2$ and $B/r_\odot^3$.

\subsection {Alternative isotach solution}\label{2pt3}

By way of comparison, consider the solution
of the thermal wind equation
that would obtain were the {\em angular momentum}
counteraligned with the entropy, rather than the angular
velocity.   (In hydrodynamic baroclinic turbulence, one
might expect the coupling to be of this nature since 
entropy and angular momentum tend to be retained by a displaced
fluid element.)  Then $S=S(l^2)$, where $l=R^2\Omega$ is
the specific angular momentum.  We denote $dS/dl^2$ by $S_{l^2}$.
Our analysis proceeds along lines identical to \S\ref{anall}, and
$l^2$ satisfies the PDE
\beq
{\dd l^2\over\dd r} -\tan\theta \left( {1\over r}
+{gr^2\sin^2\theta S_{l^2}\over C_P} \right)
{\dd l^2\over\dd\theta}=0.
\eeq
The contours of constant $l^2$ are found to be of the form
\beq\label{contsl}
{1\over R^2}={1\over r^2\sin^2\theta} = A_{l} +{B_{l}\over r}
\eeq
where $A_{l}$ is an integration constant, and now 
\beq
B_{l}= - {2GM_\odot S_{l^2}\over C_P}
\eeq
Note that $r_\odot B_{l}$ is a dimensionless constant of order unity.  
We will use this solution as a point of comparison in the next section.

\subsection {An explicit solution}\label{anexpl}

Let us turn to the data to see how our solutions fare, beginning with
our first, $\Omega$ based solution.  
As an illustrative example, consider the reduced
problem in which $S'$ is constant, not just along a particular characteristic,
but everywhere.  Then $B$ is also everywhere constant.  If $\Omega$
is now specified on some particular radial shell as a function of
$\theta$, we may write down the solution everywhere.  For this purpose,
it is easiest to choose the solar surface radius $r=r_\odot$.

Let the angular velocity at $r=r_\odot$ be $\Omega_\odot(\cos^2\theta_0)$.
We use the fit of Ulrich et al. (1988):
\beq\label{fit}
\Omega_\odot(\cos^2\theta_0)=2\pi\left( 451.5 - 65.3 \cos^2\theta_0
-66.7\cos^4\theta_0\right)\ {\rm nHz.}
\eeq
Note that $\theta_0$ carries a subscript
to indicate that it is the particular value
of $\theta$ along each trajectory characteristic (\ref{conts})
that intersects
the surface shell $r=r_\odot$.  Equation (\ref{conts}) becomes
\beq\label{cont1}
r^2\sin^2\theta = r_\odot^2\sin^2\theta_0 +B\left( {1\over r_\odot}
-{1\over r} \right),
\eeq
or
\beq\label{cont2}
\cos^2\theta_0 =  1- x^2 +
\left(B\over r_\odot^3\right) \left( 1
-{1\over \varpi} \right)
\eeq
where
\beq
\varpi=r/r_\odot,
\qquad 
x^2 = \varpi^2\sin^2\theta .
\eeq
(Equations (\ref{cont1}) and (\ref{cont2}) are also valid if 
$B$ is a function of $\cos\theta_0$.)
Substituting equation (\ref{cont2}) for $\cos^2\theta_0$
into equation (\ref{fit}) then generates the solution 
everywhere in the shell.  

\begin{figure}
\epsfig{file=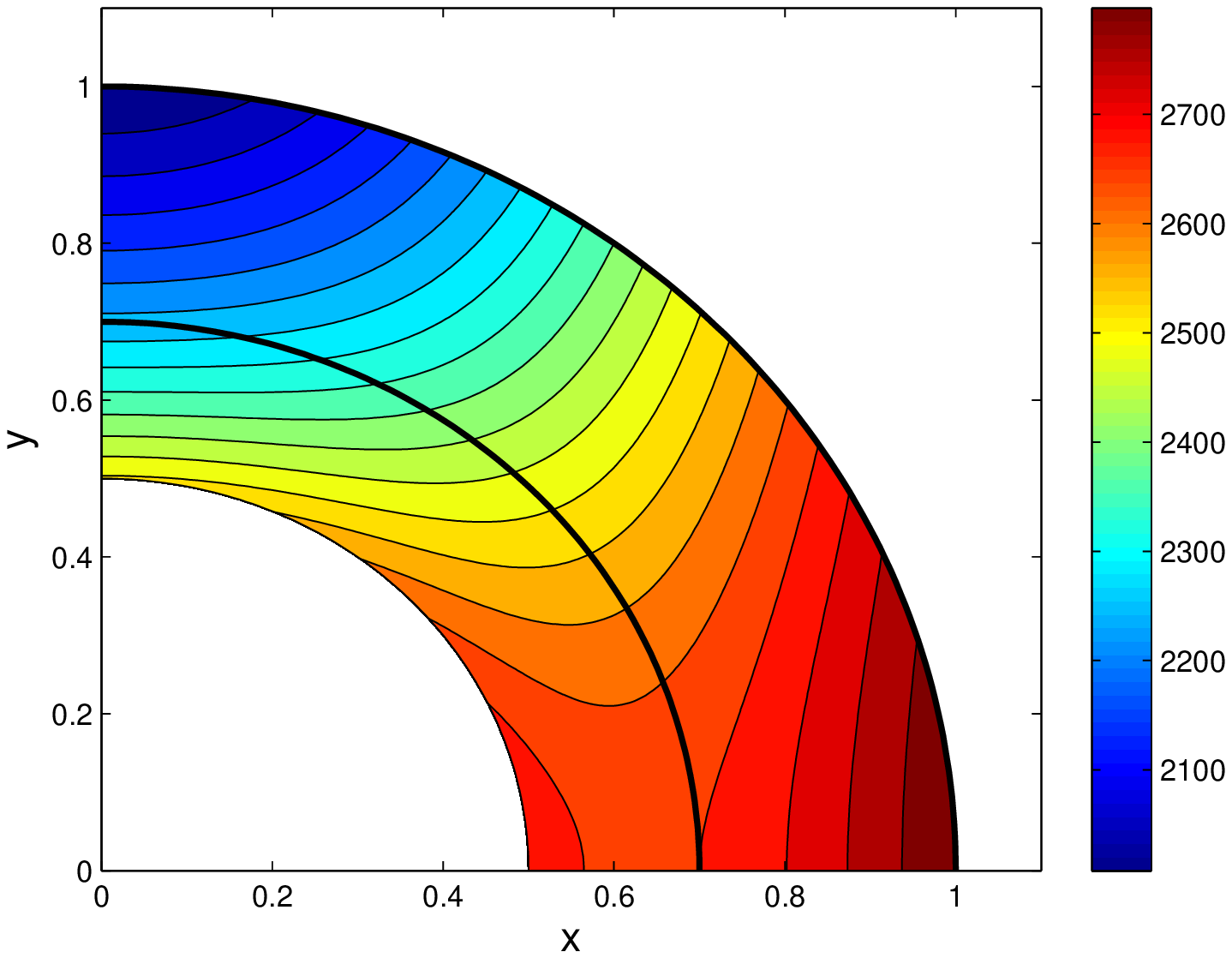, width=8 cm}
\epsfig{file=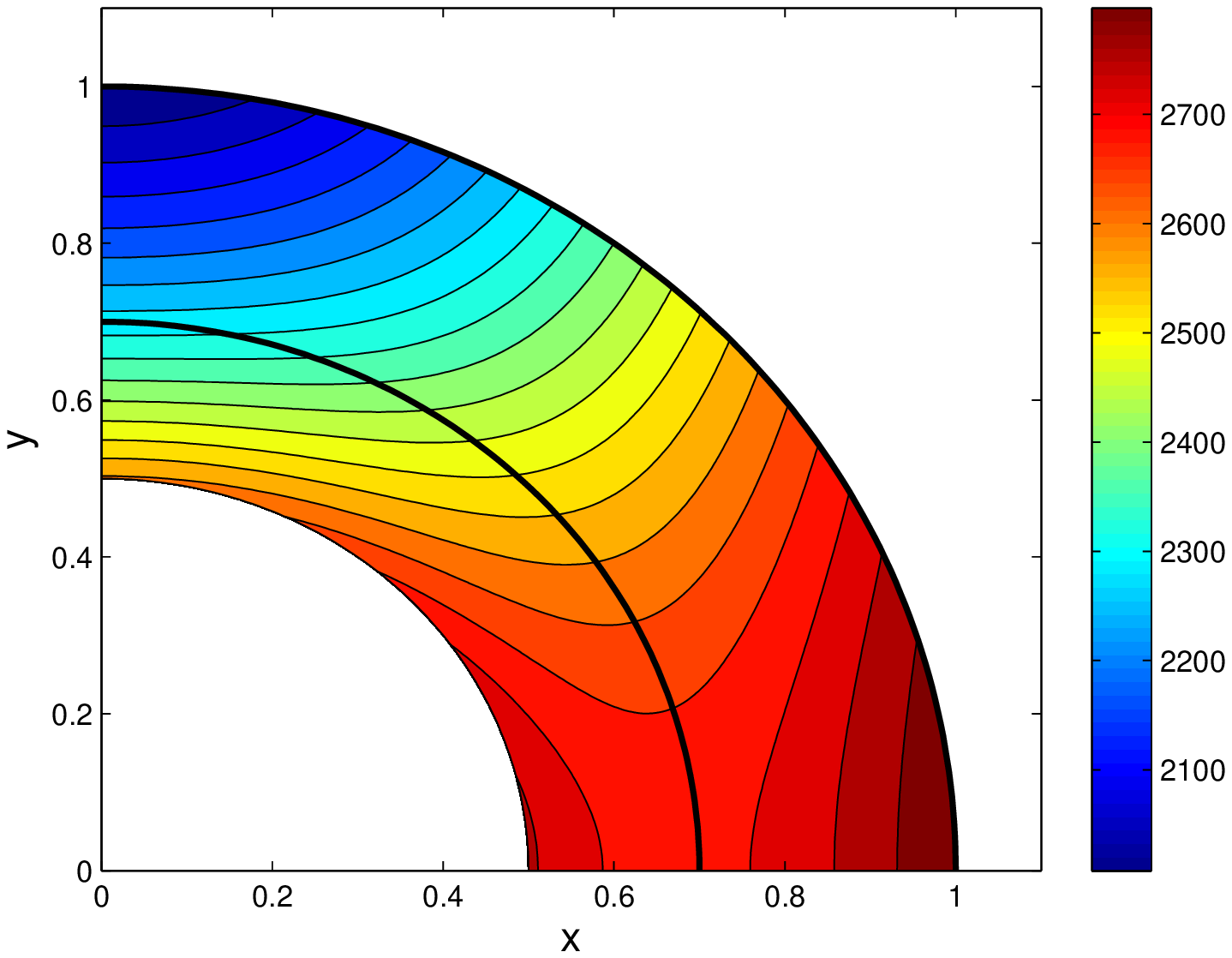, width=8 cm}
\caption {Contour plot of $\Omega(r,\theta)$ in solar interior, using
surface fit of Ulrich et al. (1988).  SCZ boundaries marked
in white.  Calculation based on
eqs. [\ref{fit}] and characteristic equations of isorotational
contours, [\ref{cont2}].  $B/r_\odot^3=0.5\,{\rm (left)},\,  0.6\, {\rm (right)}$.}
\end{figure}

The result of this procedure is shown in figure (1) for the cases
$B/r_\odot^3=0.5, 0.6$.  The detailed fit of the contours to the actual
data is not perfect---there is no tachocline in these simple models,
and the true high latitude isotachs show stronger curvature, following
spherical shells, before turning upward.   But the overall trend of the
isotachs being predominantly quasi-spherical at high latitudes, increasingly
radial at midlatitudes, and axial at small latitudes is unmistakable.
Moreover, fitting our solution to the observed surface data, while
convenient, is unlikely to show off its best form: thermal wind balance
probably breaks down near the solar surface (Thompson et al. 2003).
Given the simplicity of our direct approach, the qualitative
agreement is both striking and encouraging.

The iso-angular-momentum contours of equation (\ref{contsl})
can also be used to construct
an explicit solution.  In this case, the surface angular momentum
$l$ 
fit is
\beq\label{ll}
l(\cos^2\theta_0) =2\pi r_\odot^2 \sin^2\theta_0
\left( 451.5 - 65.3 \cos^2\theta_0
-66.7\cos^4\theta_0\right).
\eeq
Instead of equation (\ref{cont2}), we have
\beq\label{cosl}
\cos\theta_0^2 = 1-\sin^2\theta_0=
1 - \left[ {1\over x^2} +r_\odot B_{l} \left(1-{1\over
\varpi} \right) \right]^{-1}
\eeq
Substitution of (\ref{cosl}) into (\ref{ll}) generates the full solution
for the specific angular momentum $l(r,\theta)$, and the angular
velocity solution follows immediately from
\beq\label{cosll}
\Omega(\cos^2\theta_0) = (2\pi/x^2)\sin^2\theta_0
\left( 451.5 - 65.3 \cos^2\theta_0
-66.7\cos^4\theta_0\right)
\eeq

\begin{figure}
\epsfig{file=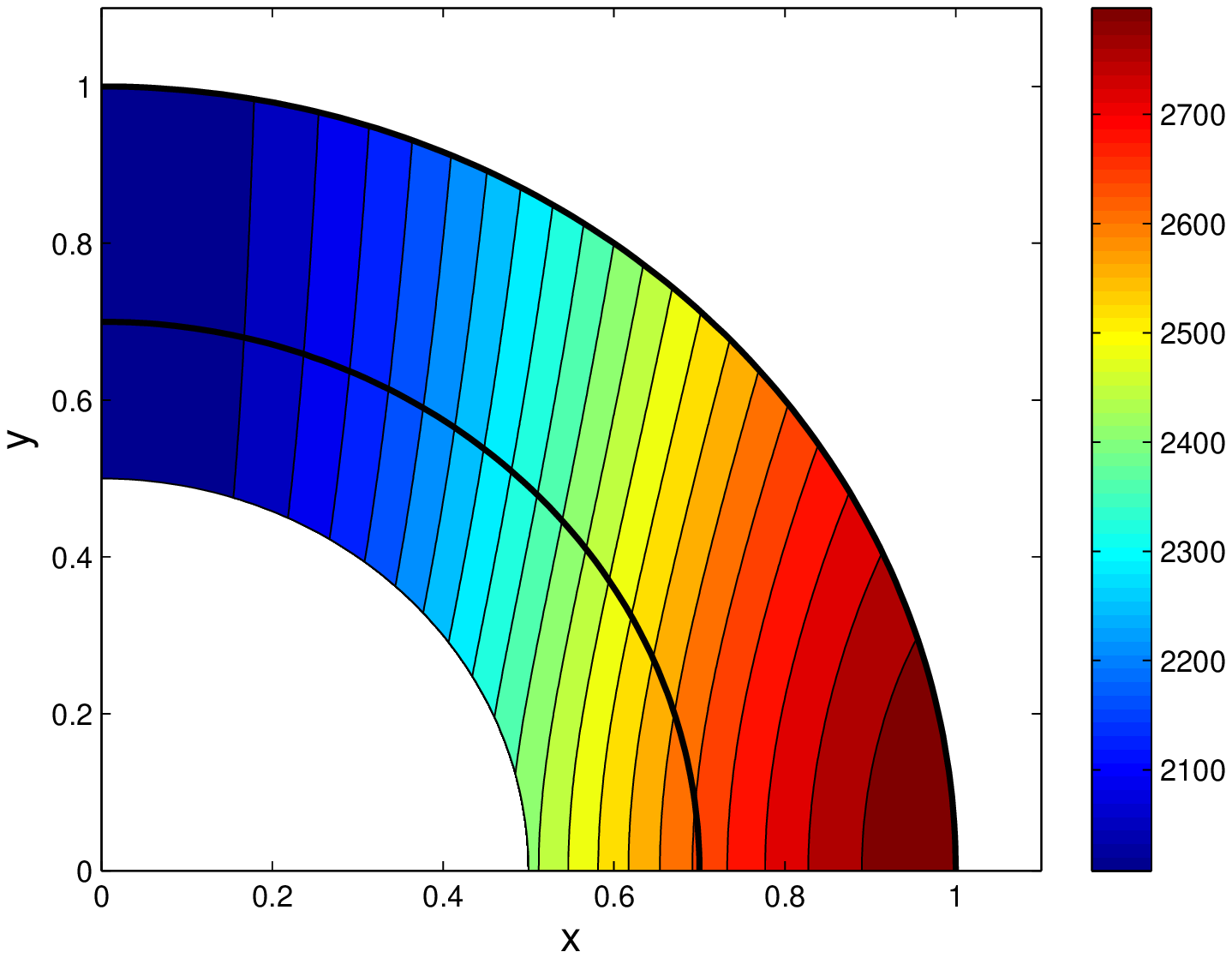, width=8 cm}
\epsfig{file=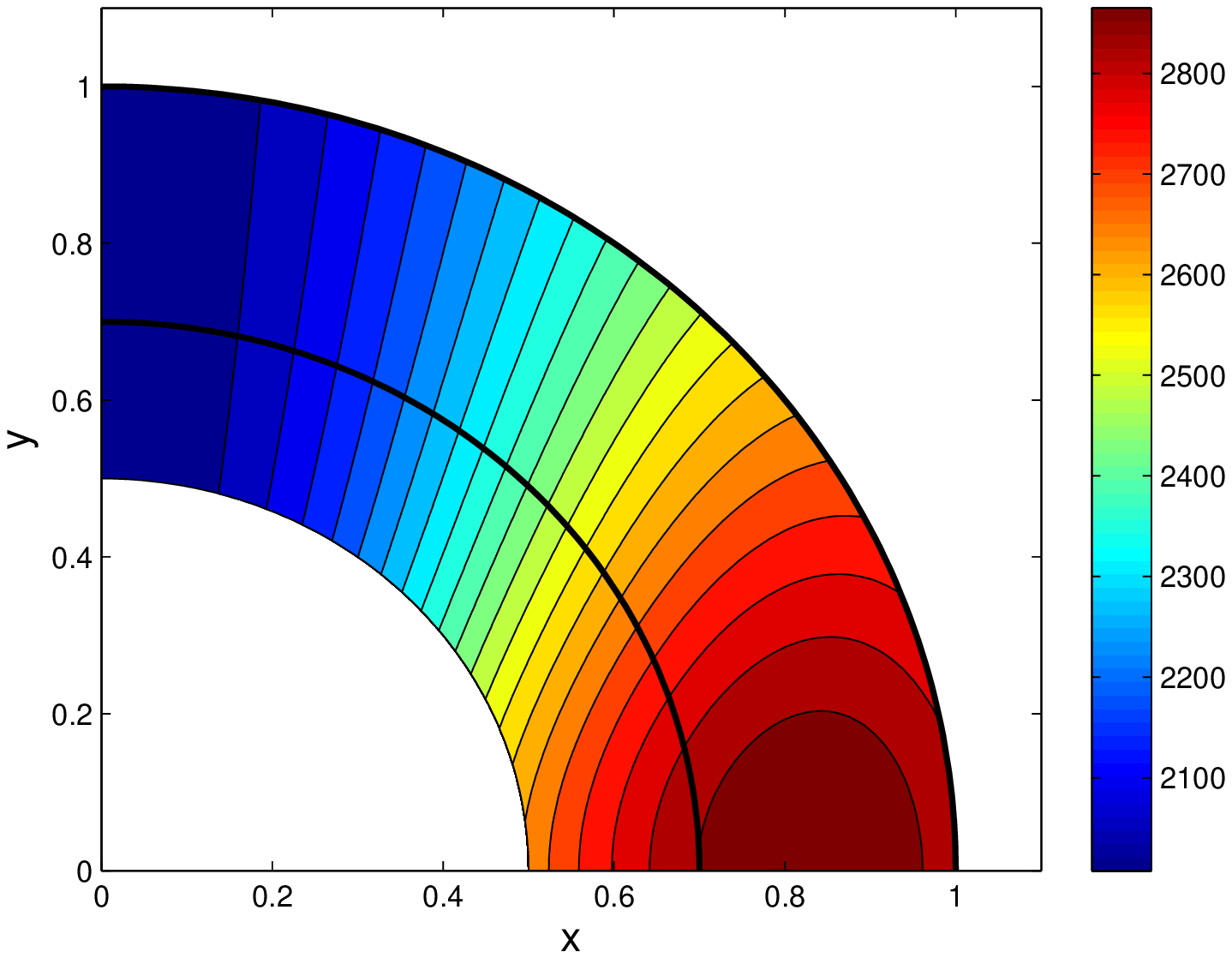, width=8 cm}
\caption {Contour plot of $\Omega(r,\theta)$ in solar interior,
based on counter aligned entropy and angular momentum gradients,
using
surface fit of Ulrich et al.\ (1988).  SCZ boundaries marked
in white.  Calculation based on
eq. [\ref{cosll}] and characteristic equation of iso-angular-momentum
contours, [\ref{ll}].
$r_\odot B_{l^2}=0.2\,{\rm (left)},\,  0.5\, {\rm (right)}$.
}
\end{figure}

In figure 2, we show two representative diagrams of the isorotational
contours taken from this alternative angular momentum based approach.
In general the contours are too cylindrical, to some extent exhibiting
the same syndrome often seen in numerical SCZ simulations.  The contrast
between figures 1 and 2 is very apparent.  There seems to be a real
linkage between $S$ and $\Omega$, and it matters very much that the
coupling is between $S$ and $\Omega$, not $S$ and $l$.  It is possible
that the refractory nature of the cylindrical contours of the simulations
is due to an $S-l$ coupling that remains too strong, as noted in in
\S \ref{2pt3}.  While it is possible that this may be cured by a more
highly resolved treatment of the turbulent fluid, it is also possible
that magnetic fields may be playing a non-negligible role, enforcing
an $S-\Omega$ coupling by field line tethering of the fluid elements.
We pursue this possibility in \S 3.

In the remainder of the paper, we focus exclusively on our original
$S-\Omega$ solution.  

\subsection {Tightening the contours}

By allowing the $B$ parameter to vary from one isotach to another,
the isorotational contours we have found
can become more tightly spaced near the poles.
In this sense, our solutions admit, but do not
demand, something reminscent of tachoclinic structure.
Equation (\ref{conts}) may be written
\beq\label{tach1}
r= {B/A\over 1 -(R^2/A)}
\eeq
If the variation of $\Omega$ leads to a nearly constant ratio
for $B/A$ close to the outer radius of the radiative zone
$r_{rad}$,
but very different $A$ values, tachoclinic structure
results.  In the polar regions near the axis of rotation,
the contours would all converge to $r_{rad}$, while                  
corresponding to very different $\Omega$ values.
This would look very much like a tachocline.

Does this make physical sense?   
Equation (\ref{cont1}) implies $A=r_\odot^2 \sin^2\theta_0
+B/r_\odot$, or
\beq
{B\over A} =  {r_\odot\over 1 + (r_\odot^3\sin^2\theta_0/B)},
\eeq
assuming that the data are initially specified on $r=r_\odot$. 
Hence, $B\propto \sin^2\theta_0$ would produce tachoclinic
structure.


Physically , this would mean that $B\propto dS/d\Omega^2$ is small
and negative near the pole, and that the entropy decreases toward
the equator as $\Omega$ increases.  In the vicinity of the equator the
entropy drops sharply.  This is not unreasonable behavior: near the pole,
unencumbered by Coriolis deviations, convection is most effective, whereas
at the equator, the opposite is true.  At the same time, the observations
suggest that $\Omega$ is changing rapidly near the pole, and only slowly
at the equator.  This is consistent with our simple picture.

We return to equation (\ref{cont2}) and
replace $B/r_\odot^3$ by
\beq\label{eta1}
B/r_\odot^3 = \eta_1 +\eta_2\sin^2\theta_0.
\eeq
Mathematically, this corresponds to the first two terms in 
a Taylor series expansion of $S'$ as a function of
$\sin^2\theta_0$
(or equivalently $\cos^2\theta_0$).
By varying $\eta_1$ and $\eta_2$ we can go between
a singular tachocline $\eta_1=0$ and the 
previous case $\eta_2=0$.  Substituting (\ref{eta1})
in (\ref{cont2}) and solving for $\cos^2\theta_0$
gives
\beq\label{tacos}
\cos^2\theta_0 = 1 - 
{\varpi x^2  -\eta_1(\varpi-1) \over
\varpi+\eta_2(\varpi-1)}.
\eeq
Using equation (\ref{tacos}) in (\ref{fit}) now produces
the interior structure $\Omega(r,\theta)$.

With two parameters available, one can of course produce somewhat
better fits to the rotation profile.  In practice, the improvement over
\S\ref{anexpl} is noticeable, but not dramatic.  In figure 3, we show on
the left the isotachs for $\eta_1=0.3$, $\eta_2=0.2$.  This gives a very
respectable fit to the data away from the tachocline, $r\ge 0.75 r_\odot$,
say.  If we wish to include an explicit tachocline in our modeling, our
earlier considerations suggest that we should restrict ourselves to small
values of $\eta_1=0$.  On the right side of figure 3, the interesting case
of $\eta=0.12$ and $\eta_2=0.8$ is presented.  A striking ``tachocline''
structure appears, though formally it lies just beneath the SCZ.  The solar
midlatitude radial contours and equatorial cylinders regions are, however,
rather well-represented.  Note as well the gentle nonmonotonic behavior of
$\Omega(R)$ near the equator, a feature seen in the helioseismology data.
Increasing the value of $eta_2$ to bring the tachocline to larger radii
(while keeping the surface layers fixed) appears to cause too much global
distortion, though an exhaustive parameter search has not been performed.

\begin{figure}
\epsfig{file=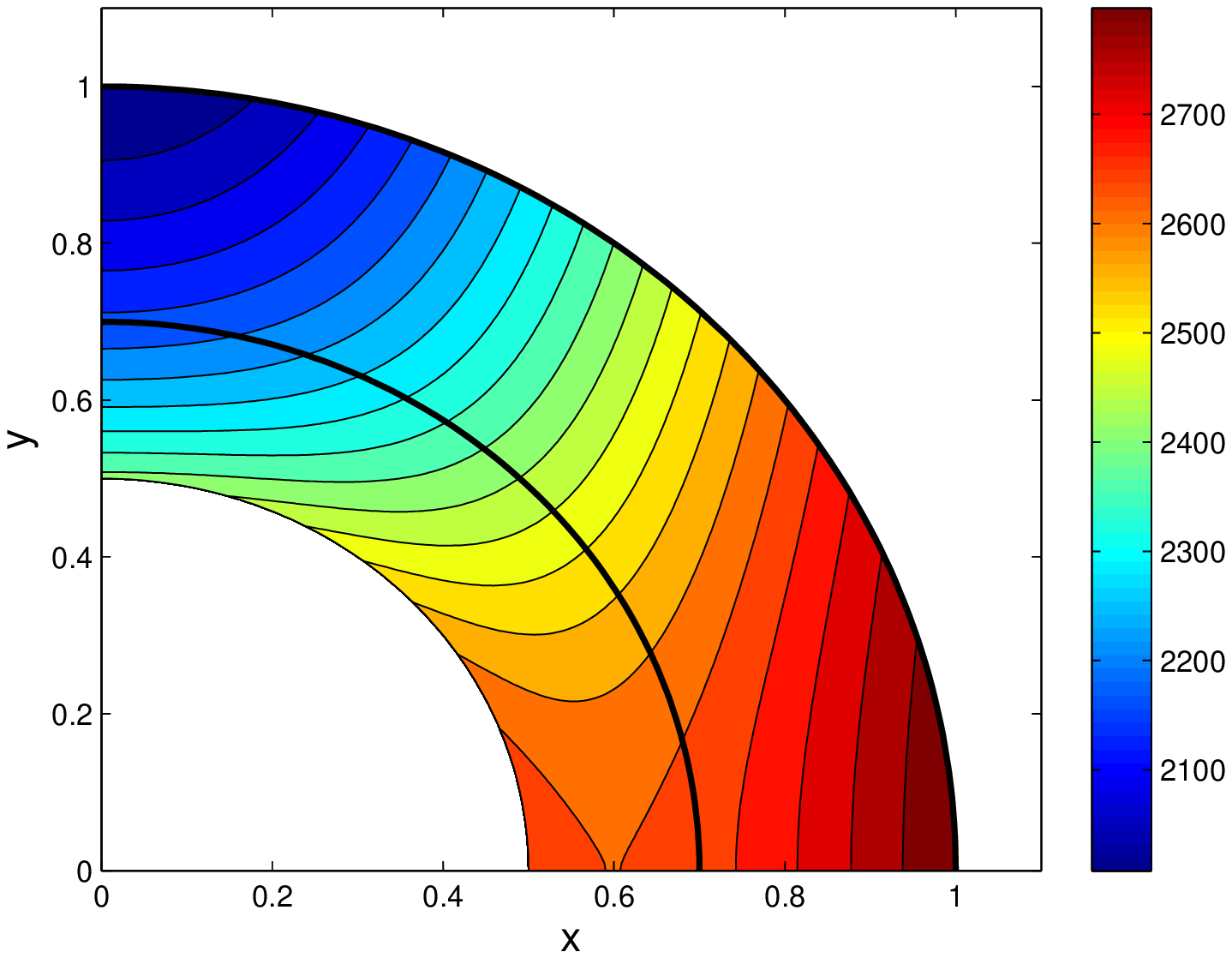, width=8 cm}
\epsfig{file=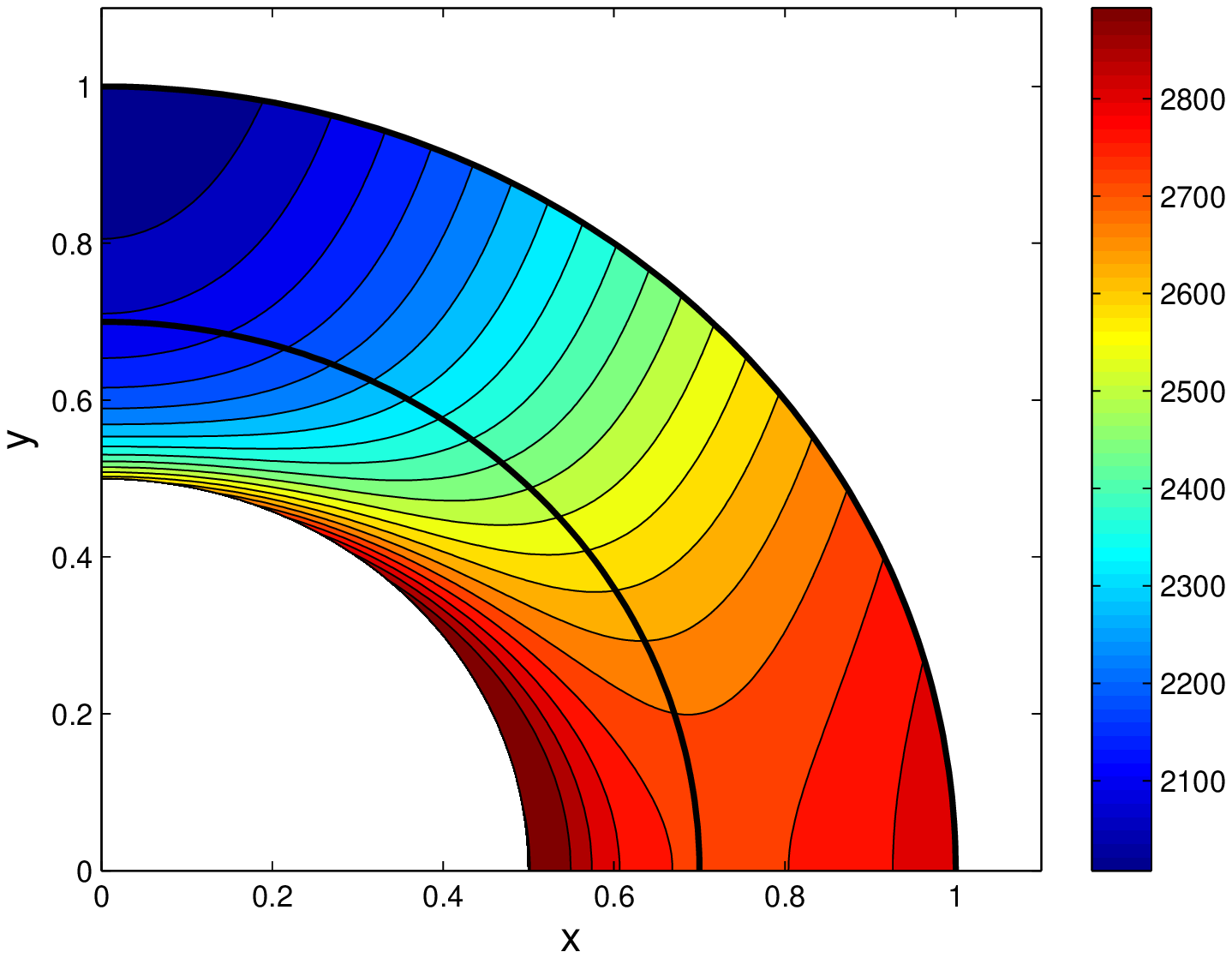, width=8 cm}
\caption 
{As in figure 1, but calculation now based on
equations [\ref{fit}] and 
[\ref{tacos}] with $\eta_1=0.3$, $\eta_2=0.2$ (left), 
and $\eta_1=0.12$, $\eta_2=0.8$ (right).  The two-parameter fit
on the left is a slight improvement over the earlier single parameter
models.  The fit on the right is striking in its
overall resemblance to the SCZ, though the ``tachocline''
formally lies
below the convective zone lower boundary.  Equatorial
and midlatitude contours are well-represented.  }
\end{figure}

At this stage, these results are suggestive, but not more than that.
It seems likely that more complex choices for $S'(\Omega^2)$
could improve the contour fits, but the agreement is impressive
even in the simplest models.  
A truly compelling explanation of the tachocline
will involve more than just the thermal wind equation and the outer
convective zone layers.  The solar tachocline arises from the complex
coupling of the rigidly rotating radiative core and the overlying strongly
shearing convective zone, not the demands of the surface rotation and
vorticity conservation, and very different dynamical processes are
likely to be involved.  Our simple, prescription valid 
above the tachocline ($r\gtsim 0.75 r_\odot$)
may represent a sort of outer SCZ solution
that asymptotically matches onto an inner solution in which tachocline
dynamics become locally dominant.

\subsection {Generic features of isotachs}

The isotachs we have found have a very distinctive ``viking helmet''
structure.  This unusual feature is also characteristic of solar
isorotation contours, and worth examining in isolation.

To understand the general structure of the isotachs,
rewrite eqation (\ref{tach1}) as
\beq
{r\over r_\odot} = {B/Ar_\odot \over 1 - (r_\odot^2/A)(R^2/r_\odot^2) }
={\alpha\over 1-\beta R^2/r_\odot^2}
\eeq
which defines our two parameters $\alpha$ and $\beta$,
\beq
\alpha=B/r_\odot A, \qquad \beta=r_\odot^2/A
\eeq
We now transform to Cartesian coordinates in a meridional plane,
\beq
y= (r/r_\odot)\cos\theta, \quad 
x= (r/r_\odot)\sin\theta.
\eeq
The Cartesian contour structure is slightly 
easier to visualize:
\beq\label{ycont}
y = \left[ {\alpha^2\over (1-\beta x^2)^2 } -x^2 \right]^{1/2}
\eeq

We must of course view equation (\ref{ycont}) through the convective
zone slot $0.7<\sqrt{x^2+y^2}<1$.  There are three types of contours.
The first is a typical polar contour, emerging perpendicular
to the axis before
bending upward, seen at high latitudes in figure 1.  The second class,
visible at midlatitudes in figure 1, is hidden in the radiative zone
at small $x$, and
does not emerge into the SCZ until the contour
is well separated from the axis, at which point
the isotach has a quasi-radial character.  The third contour class is deeply
buried, running along a very small radius in the core (not seen),
completely disappearing in the bulk of the sun ($y$ is imaginary).
It
then makes a sudden leap upward into the equatorial region of the SCZ,
where it appears nearly axial.  These are the low latitude regions of figure
1, corresponding physically to Taylor columns of constant rotation
on cylinders.  Equation (\ref{ycont}) thus displays the three key
traits one associates with isotachs: horizontal near the poles, radial
at midlatitudes, and cylindrical near the equator.

\section{Axisymmetric modes in a rotating, baroclinic, weakly magnetized
gas}

\subsection {Preliminaries}

The starting point of the analysis of \S 2 was that the dominant balance
of the vorticity equation is given by the thermal wind relation.
We adopted a phenomenological connection between $S$ and $\Omega$
that succeded in reproducing the observed general behavior of the solar
isorotational contours.  In this section we examine a possible reason for
this coupling.  We suggest that the underlying cause of the $S-\Omega$
coupling is to be found in the general dynamical stability properties
of a magnetobaroclinic fluid.

We have already noted that in accretion disk applications, the combination
of magnetic fields and differential rotation is extremely destabilizing,
even if the field is very weak and not particularly well-ordered.  This is
a consequence of the MRI.  The analysis presented here is an extension
of the accretion disk problem, but a highly significant extension: we
consider the most general dynamical axisymmetric response to a weakly
magnetized baroclinic system.  To avoid possible confusion, we will
reserve the term ``MRI'' to apply only to the magnetic destabilization
process in rotation-dominated disks.  The topic of this section is
the ``magnetobaroclinic instability.''

The system we analyze is a proxy that shares important features 
with the sun.  It consists of a body of self-gravitating gas
that has arbitrary axisymmetric angular velocity and entropy profiles.
Instability in the form of turbulent thermal convection is treated on the
same footing as rotational instability.  Both the thermal and rotational
profiles can in principle be altered by turbulent fluxes arising from
magnetobaroclinic instability, since neither profile is fixed by the
requirements of hydrostatic equilibrium.  This differs from the behavior
of an accretion disk, whose rotational profile, generally Keplerian,
is not at liberty to change.

It is an elementary fact that a convectively unstable stratified gas
tends to alter its thermal gradient to a nearly adiabatic configuration,
thereby regulating the linear instability itself.  What is novel here is
that we extend this notion to include simultaneously both the rotational
and thermal responses.  This is generally not something investigators have
pursued, because at first sight it does not appear to be particularly
promising.  Hydrodynamically, the differential rotation of the sun is
not close to instability.  It is only when magnetic fields are considered
that rotational instabilities are raised to the same level as convective
instabilities.  Indeed, even a uniformly rotating magnetized gas is
just ``marginally stable,'' tipping to instability with only a slightly
adverse angular velocity gradient.

The invocation of magnetic instability may strike some readers as
dubious.  Is a turbulent convective zone fertile ground for process
that requires at least some degree of field coherence?  Is it justified
in the analysis to prescribe $\Omega$ {\em a priori} when in fact it
is built up by convection?  These questions can ultimately be settled
only by well-designed numerical simulations.  In the meantime, let us
first understand the behavior of our proxy magnetic system, a challenge
in itself.  We will then be in a better position to address more thorny
issues.

\subsection {Stability of a magnetized baroclinic gas}\label{mbcg}

We seek to understand the linear stability properties of a gas in which convective,
rotational, and magnetic effects are treated as co-equals.
As before, let $(R, \phi, z)$ be a standard cylindrical coordinate system.
Consider an axisymmetric rotating gaseous body
whose equilibrium angular velocity $\Omega$ and
entropy $S$ are 
allowed to be functions of both $R$ and $z$.
We consider local perturbations of plane wave form,
$\exp(i\bb{k\cdot r} -i\omega t)$.  Here, $\bb{k}$ is the
local axisymmetric wavenumber, $\bb{r}$ the position vector, $\omega$
the angular frequency (or growth rate) and $t$ is the time.
Such disturbances satisfy the 
dispersion relation (Balbus 1995, Balbus \& Hawley 1998):
\beq\label{disprel1}
{k^2\over k_z^2}\varpi^4
+\varpi^2\left[ {1\over\gamma\rho}({\cal D} P){\cal D}(\ln P\rho^{-\gamma})
+{{\cal D}(R^4\Omega^2)\over R^3} \right]
-4\Omega^2(\bb{k\bcdot v_A})^2
= 0
\eeq
where $\bb{v_A}$ is the Alfven velocity
$$
\bb{v_A}= {\bb{B}\over\sqrt{4\pi\rho}}
$$
and
\beq
{\cal D} \equiv \left( {k_R\over k_z}{\dd\ \over\dd z} - {\dd\ \over\dd R}\right),
\quad
\varpi^2=\omega^2-(\bb{k\bcdot v_A})^2.
\eeq

We follow the stability arguments of Balbus (1995).  
The variable
$\varpi^2$, and hence $\omega^2$,
must be real.  We may therefore determine stability by noting those
conditions under which $\omega^2$ passes through zero.  
The solution $\omega^2=0$ is possible if
\beq
(\bb{k\bcdot v_A})^2 = {k_z^2\over k^2}\left(
4\Omega^2 +{1\over\gamma\rho}({\cal D} P){\cal D}(\ln P\rho^{-\gamma})
+{1\over R^3} {\cal D}(R^4\Omega^2)\right).
\eeq
To assure stability, this equation cannot have any solutions
for $(\bb{k\bcdot v_A})^2$, hence the right side must satisfy
\beq\label{magstab}
4\Omega^2 +{1\over\gamma\rho}({\cal D} P){\cal D}(\ln P\rho^{-\gamma})
+{1\over R^3} {\cal D}(R^4\Omega^2)<0,
\eeq
a condition that does not involve the magnetic field, though it
pertains {\em only} to a magnetized fluid!  Notice in particular
that the field geometry is unimportant.  The hydrodynamical
stability condition, by way of contrast, would be
\beq\label{hystab}
{1\over\gamma\rho}({\cal D} P){\cal D}(\ln P\rho^{-\gamma})
+{1\over R^3} {\cal D}(R^4\Omega^2)<0 \quad {\rm (hydrodynamic\ stability)},
\eeq
an altogether different and far more easily satisfied requirement.  

If, in equation (\ref{magstab}),
we set $x=k_R/k_z$ and expand the ${\cal D}$ operator, we may recast
the inequality as
\beq\label{xpoly}
x^2 N_z^2 + x\left[ {1\over\gamma\rho} \left(
{\dd P\over \dd z}{\dd\ln P\rho^{-\gamma}\over \dd R} +
{\dd P\over \dd R}{\dd\ln P\rho^{-\gamma}\over \dd z} 
\right) -R{\dd\Omega^2\over \dd z}\right]
+N_R^2 +{\dd\Omega^2\over \dd\ln R} >0,
\eeq
where
\beq
N_z^2 = -{1\over\rho\gamma}{\dd P\over \dd z}{\dd \ln P\rho^{-\gamma}
\over \dd z}, \qquad 
N_R^2 = -{1\over\rho\gamma}{\dd P\over \dd R}{\dd \ln P\rho^{-\gamma}
\over \dd R}.
\eeq
In the thermal wind models of \S 2, $N_z^2 >0$, $N_R^2<0$,
and $N_R^2 +{\dd\Omega^2/ \dd\ln R} >0$
throughout the bulk of the SCZ at midlatitudes.  

Two conditions will ensure that the quadratic polynomial in $x$ is positive.
We refer to these as the magnetized H{\o}iland criteria, after the
investigator who solved the corresponding hydrodynamic problem
(Tassoul 1978).  The first is
\beq\label{hoi1}
N_R^2 + N_z^2+ {\dd\Omega^2\over \dd\ln R}
= N^2+ {\dd\Omega^2\over \dd\ln R} >0,
\eeq
since this means that either very large or very small $x$ is positive.
This criterion generally seems to be satisfied throughout the bulk of
the convection zone.  The second criterion follows from requiring that
the discriminant of the quadratic $x$ polynomial (\ref{xpoly}) should
be negative so that there are no real roots.
The result of this
somewhat lengthy calculation is\footnote{This result was first derived using a variational 
approach by Papaloizou \& Szuszkiewicz (1992).}
\beq\label{hoi2}
\left( - {\dd P\over\dd z} \right)
\left(
{\dd\Omega^2 \over \dd R}{\dd\ln P\rho^{-\gamma}\over \dd z}-
{\dd\Omega^2 \over \dd z}{\dd\ln P\rho^{-\gamma}\over \dd R}
\right) >0.
\eeq
In the course of deriving this result, we explicitly use
the $\phi$ component of the vorticity equation,
\beq\label{vort1}
R {\dd\Omega^2\over \dd z} = {1\over\rho^2}
\left( {\dd\rho\over\dd R}{\dd P\over \dd z} -
{\dd\rho\over\dd z}{\dd P\over \dd R} 
\right),
\eeq
which will  be recognized as the starting point for our thermal 
wind analysis in \S 2.   It can be shown that equations (\ref{hoi1})
and (\ref{hoi2}) together imply
\beq
N_R^2 + {\dd\Omega^2\over \dd\ln R} >0, \qquad N_z^2 >0
\eeq
a somewhat more stringent requirement than equation (\ref{hoi1})
by itself, but an obvious set of constraints that also follows
simply upon inspection of equation (\ref{xpoly}).    

The second magnetized H{\o}iland criterion (\ref{hoi2}) states that the
$\phi$ component of $\bb{\nabla} S \bb{\times\nabla} \Omega$ should be positive
in the northern hemisphere and negative in the southern hemisphere for
stability.  (See figure 4.)  Marginal stability by this second important
criterion, would correspond to entropy and angular velocity surfaces
coinciding.  In our problem, the gradients would be oppositely directed.
It is to be noted that this criterion holds regardless of field geometry
or strength, as long as the Alfv\'en velocity is relatively small.

\begin{figure}
\epsfig{file=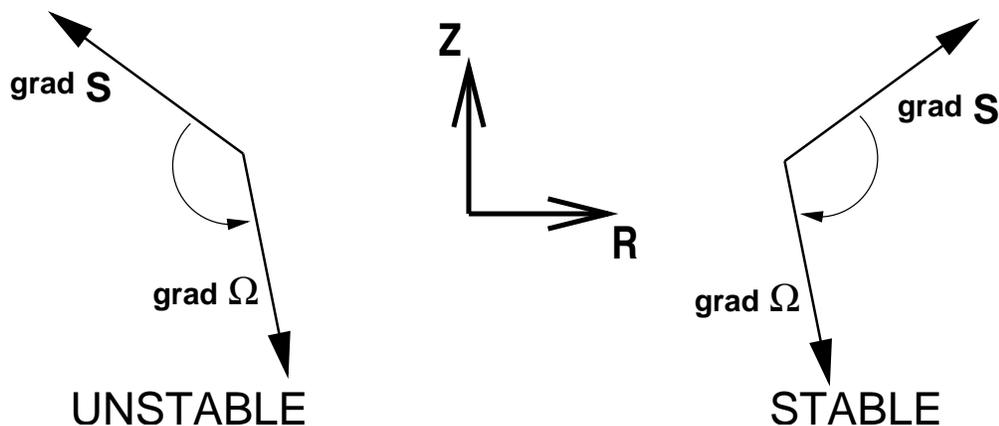}
\caption 
{An UNSTABLE (left) and a STABLE (right) alignment
of $\del S$ and $\del \Omega$ for northern hemisphere
disturbances.  Marginal stability,
used in our solution of the thermal wind equation (\ref{thw2}) 
corresponds to precise counter-alignment of these gradients.}
\end{figure}

The marginalization of the linear instability must be viewed
at present as a plausible outcome of the induced turbulent flow,
rather than a certainty.  But assuming that the system ultimately does
arrange its entropy and angular velocity gradients to curtail unstable
magnetobarotropic modes, it would appear to do so by passing through
gradually diminishing values of $(\bb{k\cdot v_A})^2$.  Eventually the
wavelength will exceed the size of the physical domain, and there can
be no question of an unstable mode at this point.  In practice however,
the relevant lengthscale is likely to be determined by the coherence
length of the magnetic field, not the size of the SCZ.  This is probably
a dimension not very different from the largest convective eddy scale.

\subsection{Marginal magnetobaroclinic instability}

Section \ref{mbcg} suggests that the underlying dynamical explanation of
the near coincidence of constant entropy and angular velocity surfaces,
which seems to be a good phenomenological model of the helioseismology
data, is marginal stability to axisymmetric magnetobaroclinic modes.
This raises a number of questions.  Why these modes?  Why focus on
axisymmetry?  Might nonaxisymmetric modes be more unstable?

The answer to the first question is that the modes considered are just
the standard convective motions at the heart of solar turbulence,
but which find themselves subject to magnetic fields and rotation.
Marginal instability arguments and near adiabatic temperature profiles
for stellar convection zones are uncontroversial.  They are based on
simple physical reasoning, not complex turbulence calculations.  Here we
suggest that there is an important augmentation to these arguments
needed when both magnetic fields and rotation are present.  Because of
the coupling introduced by the magnetic field, rotational instability
must be considered with convection from the very start.

To see why this might be so, as well as to answer the second and third questions,
begin with the {\em nonaxisymmetric} dispersion relation for
a magnetized, uniformly rotating gas.  This may be derived using exactly
the same procedure followed in Balbus (1995).  For a wavenumber
with $\phi$ component $m/R$, the dispersion relation (\ref{disprel1})
becomes
\beq\label{disprel2}
{k^2\over k_z^2}
\varpi^4
+\varpi^2\left[
{1\over\gamma\rho}({\cal D} P){\cal D}(\ln P\rho^{-\gamma})
-4\Omega^2 - {m^2\over k_z^2R^2}N^2 \right]
- 4\Omega^2(\bb{k\bcdot v_A})^2
= 0
\eeq
where 
\beq
\varpi^2 = \omega^2 - (\bb{k\cdot v_A})^2, \qquad
N^2 = N_R^2 +N_z^2 = - {1\over\rho\gamma}{\dd P\over \dd r}{\dd\ln P\rho^{-\gamma}\over 
\dd r}
\eeq
and $r$ is the spherical radius.  
Consider first a purely hydrodynamical rotating system, $\bb{v_A}=0$.
For stability, 
\beq
N^2 >0, \qquad 
({\cal D} P){\cal D}(\ln P\rho^{-\gamma})
-4\Omega^2 <0         
\eeq
The first inequality, a requirement for nonaxisymmetric modes, is the
standard Schwarzschild criterion.  Ultimately a convectively system is
characterized by a small negative value of $N^2$, just enough to maintain
marginal levels of turbulence.
The second inequality, required of all axisymmetric modes, may be written
\beq\label{xpoly2}
x^2 N_z^2 + x\left[ {1\over\gamma\rho} \left(
{\dd P\over \dd z}{\dd\ln P\rho^{-\gamma}\over \dd R} +
{\dd P\over \dd R}{\dd\ln P\rho^{-\gamma}\over \dd z} 
\right) \right]
+N_R^2 +4\Omega^2                   >0.
\eeq
This is guaranteed if
\beq
N_R^2 + 4\Omega^2 > 0, \qquad
N_z^2 >0
\eeq
These are slightly more restrictive than the nonaxisymmetric requirement,
but hardly constrain the rotation at all.  

With the inclusion of even a weak magnetic field, there is a significant
change in the stability of axisymmetric modes.  The stability requirements
of the dispersion relation (\ref{disprel2}) for any finite $\bb{k\cdot
v_A}$ are
\beq
N_R^2>0, \quad N_z^2 >0.
\eeq
The nonaxisymmetric requirement $N^2>0$ is superfluous.  The presence of a
magnetic field precludes any rotational stabilization.  

Finally, when we allow the combination of magnetic fields and differential
rotation to be present the axisymmetric modes are elevated to the role of
key players.  Now axisymmetric modes that would be stable by the Rayleigh
criterion can be destabilized.  Shearing nonaxisymmetric disturbances
still couple to convective motion, but marginalizing the \BV growth rate
is not enough to ensure dynamical stability.  The rotation profile built
up by convection must be marginalized as well (cf. equation [\ref{hoi2}]),
and this is a much more stringent requirement in weak field MHD than
in hydrodynamics.  This is the central point of this section.

\section {Discussion and summary}

Our principle conclusion---that the rotational profile of the sun (and
presumably other late type stars) is a magnetic phenomenon---is far
reaching, and many readers may still be skeptical.  Perhaps the most
controversial points are the implicit assumption that the Alfv\'enic
$\bb{k\cdot v_A}$ coupling remains vigorous on small scales in a turbulent
fluid, and that the gradients of entropy and angular velocity should be
accorded equal respect in gauging the overall dynamical stability of the
SCZ plasma.  These issues must ultimately be established or refuted by
well-designed numerical MHD simulations.  It bears emphasis, however,
that there is a precedent of nearly two decades of intensive numerical
simulations of the magnetorotational instability in astrophysical disks.
Global MHD simulations are vigorously unstable in the face of fully
developed turbulence and highly convoluted flow.  The bulk of the
magnetic energy is generally to be found at large scales in a turbulent
fluid (e.g. Fromang \& Papaloizou 2007), and the small wavelengths that
drive instability are well-coupled to the embedded fluid magnetic field.
The question is whether this is equally true of an MHD turbulent fluid
whose primary instability is convective, with the rotation profile
built up by the convection itself.  This important question remains
open, at least for the moment.

In any case, the results of \S 3 of this paper suggest that the entropy
and angular velocity in the SCZ know about each other one way or another.
Because of the properties of axisymmetric instabilities in a weakly
ionized gas, there is a dynamical basis for the belief that they may
well be functionally related: important axisymmetric instabilities
are controled when the entropy and angular velocity are counteraligned.
But the fit of the thermal wind equation solution with the helioseismology
angular velocity contours speaks for itself.  It is, at the very least,
empirical evidence for $S\simeq S(\Omega^2)$ throughout much of the SCZ.

Of course the fit is not completely perfect, nor should it be.  The actual
helioseismology data show polar contours hugging spherical shells
before turning sharply outwards (Thompson et al. 2003).  Our thermal
wind contours are more smooth and less spherically curved.  The data show
contours closing near the surface at equatorial latitudes, something that
our characteristic-based theory does not reproduce.  It is interesting
as well to note that the the marginal stability arguments break down
in this region, since $\dd P/\dd z$ approaches zero and there is no
longer a requirement of parallel entropy and angular velocity gradients.
Moreover, Thompson et al. (2003) have pointed out that thermal wind
balance appears not to hold near in the outermost layers of the SCZ,
where turbulent transport can no longer be neglected.  In fact, our
model results appear to be too smooth here, in comparison with the data.

But our approach is very simple, and that it succeeds as well as it does
is striking.  If the magnetobaroclinic marginal stability arguments of
this paper are not correct, then it is a coincidence that such arguments
lead to solutions of the thermal wind equation in broad agreement with
the helioseismology data.  If the arguments are correct, however,
then hydrodynamical simulations are unlikely to reproduce the solar
angular velocity contours, unless there is an as yet unknown purely
hydrodynamical $S-\Omega$ coupling.   In the simulations of Miesch et
al. (2006), when such a connection was put in ``by hand'' the contour
fit noticeably improved.  To reproduce the observed SCZ isotachs in
numerical simulations without such forcing, the presence of a magnetic
field may be essential, and care should be taken to ensure that the most
rapidily growing magnetobaroclionic local instabilities are resolved.
Internal dynamics should take care of the rest: the system will evolve
to counteract the instabilities to the extent it can.  An excellent
way for the flow to do this is to counteralign its entropy and angular
velocity gradients.  The resulting isotachs should then be in broad
agreement with the helioseismology data.

Capturing the local unstable modes may not be an easy task.  To do so,
wavenumbers $k$ satisfying
\beq
(kv_A)^2 \simeq  R{\dd\Omega^2\over \dd z} 
\eeq
will need to be resolved (Balbus \& Hawley 1998).  This leads to
$\lambda/r_\odot \sim 10^{-3}$ for typical SCZ values.  Since a 
fraction of this wavelength must be represented on the grid (or higher
order spectral basis functions), this is beyond the resolution
of most of the simulations performed to date.  Numericists may wish to
consider artificially enhancing magnetic effects to drive the system
toward MHD marginal stability as a possible means to improve the fit of
the computed rotation contours.

In principle, 
one may also test our claims by using the helioseismology data for
$\Omega(r,\theta)$ and working backwards.
From equation (\ref{maineq}), the quantity
\beq
r^4\sin\theta\, \cos\theta \left(\dd\Omega^2/\dd r\right)
\left(\dd\Omega^2/\dd \theta \right)^{-1} -r^3\sin^2\theta ,
\eeq
which is directly proportional to $S'(\Omega^2)$, should have the same
isocontours as $\Omega$ itself.  But to test this would require a reliable
extraction of the partial derivatives of $\Omega^2$.

The search for an explanation of the SCZ isorotational contours has
been long and not without frustration.  If the work presented here is
correct in its basic essentials, it would be a step forward.  But if
for some reason the consistency between the solutions of
the thermal wind equation and the marginal stability requirements
of local magnetobaroclinic modes is simply an accident, if weak field
instabilities are ultimately not effective in the SCZ, even this
refutation would represent a form of progress by ruling out a viable
alternative.  Regulation of the SCZ turbulence by marginal stability to
magnetobaroclinic modes is a well-posed, directly testable concept,
involving a dynamical domain that is as yet underexplored.  It merits
serious consideration.


\section*{Acknowledgments}

I am grateful to several colleagues for encouragement over the course of
this work, particularly J. Cho, E. Dormy, H. Latter, and P. Lesaffre.
E. Quataert, M. McIntyre, J. Stone, and N. Weiss offered important
constructive advice on an earlier version of this paper, as did an
anonymous referee.  I would also like to thank P. Lesaffre and H. Latter
for their aid and skill in preparing the contour figures.  This work has
been supported by a grant from the Conseil R\'egional de l'Ile de France.

\end{document}